\documentclass{article}
\usepackage{graphicx}
\begin{document}
\title{Non-hermitean delocalization in an array of wells 
with variable-range widths}

\author{V.G.Benza\\
Facolta' di Scienze, Universita' dell'Insubria\\
Via Lucini 3, 22100 Como\\
also INFM, unita' di Milano,\\ 
Via Celoria 16,20133 Milano} 
\maketitle
\begin{abstract}
Nonhermitean hamiltonians of convection-diffusion type occur 
in the description of vortex motion in the presence of a tilted
magnetic field as well as in models of driven population dynamics.
We study such hamiltonians in the case of rectangular barriers 
of variable size.
We determine Lyapunov exponent and wavenumber of the eigenfunctions
within an adiabatic approach, allowing to reduce the original
d=2 phase space to a d=1 attractor.
\end{abstract}
\section{Introduction}
Quantum mechanics in an imaginary magnetic field (IMF) 
has been studied 
by various authors, after Hatano and Nelson \cite{HN1} pointed out 
its relevance in
describing the competition between drift and pinning for a vortex
in the presence of columnar defects.
The kinetic part of the hamiltonian with IMF is a 
convection-diffusion operator; operators of this type
occur as well
in models of driven population 
dynamics \cite{MW}, \cite{Mu}, \cite{NS}.
Due to the convection term, parity is violated and
the wave function is exponentially magnified 
along a given direction; as a consequence, 
 it can keep a localized character provided its decay, 
generated by the disordered potential, wins over the IMF
amplification.

In the zero-IMF reference hamiltonian 
the decay, if disorder does not violate parity,
is characterized by  
the inverse of a single localization length on both sides.
With convection, 
when the decay is exactly compensated by the IMF magnification factor,
the wave function becomes delocalized. 
The spectrum at the critical point bifurcates, and is distributed along
a d=1  curve in the 
complex plane.
This scenario was proved to be valid 
with various potential distributions
(gaussian ,  box, Cauchy);
it has been confirmed for 
lattice models with disordered hopping
\cite{HN2},\cite{HN3},
\cite{BSB}, \cite{BZ},
 \cite{FZ}, \cite{GK}.

In the  hermitean case,  
one can turn the eigenvalue problem into a Langevin equation, 
and determine the spectral properties \cite{BIH}
from the 
steady state distribution of the Langevin problem.

Here we generalize this approach  
to complex-valued stochastic equations,
as needed when dealing with non-hermitean hamiltonians.
Rather than starting
with a gaussian $\delta$-correlated potential, which would lead 
to a Langevin equation,    
we consider a piecewise constant potential $V(x)$,
a model 
originally studied 
by Benderskii and Pastur \cite{BP} in the hermitean case.
 The reasons
for our choice will become clear in the sequel.
If, e.g., $V(x)={V_{0},\,V_{1}}$ 
with $V_{0}=0$ and
$V_{1}<0$, $V(x)$ describes equally engineered  trapping defects 
in an homogeneous medium.

 We assume 
that  free segments and trapping segments  are independent random variables,
with  fixed 
average length $l_{0},\,l_{1}$ respectively. 
Although the stochastic equation is no longer of 
Langevin type,  the 
probability distribution still satisfies a differential equation.
This equation allows for an
 adiabatic approach, which would be impossible with white noise. 
We can thus reduce the phase space to a one-dimensional 
attractor $\mathcal{L}$, strongly simplifying the search for steady
state.
  
 The ``dynamic variable''  of the stochastic equation 
 is the
 logarithmic 
derivative $z(x)$ of the wave function. 
From the mean value  $<z(E)>,\,\,(E = E_{r}+i \cdot E_{i})$
one extracts   the inverse of the localization length, as well as 
a wavenumber.
 Generically we find that the localization length
decreases
when $E$ departs from the real axis, and  increases with $E_{r}$.

This behavior implies that,
in the presence of an IMF,
 at a critical value of $E_{r}$ the system delocalizes;
the contour lines of  $<z_{r}(E)>$ determine then the  
mobility edge in the complex energy plane.
Three different  regimes (low, high and intermediate $E_{r}$) will be 
analyzed in the next three sections.
 
\section{Langevin formulation of the quantum eigenvalue problem}
The eigenvalue problem ${\hat H} \psi = E \cdot \psi$ for a d=1 Hamiltonian 
${\hat H}_{0} = - {d^2 \over dx^2} + V(x)$ with disordered potential $V(x)$ 
translates into a Langevin equation for the logarithmic derivative $z$ of
the wave function $\psi$:
\begin{equation}
\label{A1}
{dz \over dx} = -(z^2 + E) + V(x);\,\,\,z={ \psi^{'} \over \psi}.\\
\end{equation}
 As 
 is well known, 
 one can limit the analysis to real-valued wave functions.
The deterministic part of the equation is invariant under 
space inversion $(z(-x)=-z(x))$ ; hence, 
if the statistical properties of $V(x)$ are also invariant, 
the backward and forward  $x$-evolution of $z$ 
are equivalent, i.e. they approach the same steady state. 
One  defines the Lyapunov exponent  $\gamma(E)$:
\begin{eqnarray}
\gamma(E)&=&lim_{x \to \infty} {<ln r(x)> \over x}\nonumber\\
r^{2}(x)&=&\psi^{2}(x) + \psi'^{2}(x)\nonumber
\end{eqnarray}
where the average is over  $V(x)$; 
$ \gamma(E)$ is always positive  for disordered
potentials with rapidly decaying correlations
\cite{LGP}.
The following identity is shown to hold:
\begin{equation}
\label{Lya}
\gamma(E)= <z>\nonumber 
\end{equation}
The Lyapunov exponent is an index for the exponential growth rate 
of the wave function $(\psi(x) \approx exp[\pm \gamma \cdot x]\,\,
x \to + \infty)$: here, by Oseledec's theorem,  the plus sign
always occurs, but for  a single case, corresponding to the
physical (square-summable) solution \cite{BS}.
Consistently, one has indeed $<z>\,>0$ and 
a non-even steady state 
distribution $P(z)$.

Let us now examine how this picture is changed upon  
adding an external (constant) imaginary  magnetic 
field (IMF). In the hamiltonian, this is mimicked by:
${\hat H}= -({d \over dx} - a)^{2} + V(x)$, 
 implying  $ z \to z-a$ in Eq. \ref{A1}, which now is to 
be studied for complex $z$ and $E$ ( $z=z_{r} + i \cdot z_{i}$,
 $E=E_{r} + i \cdot E_{i}$).
 The space-inversion invariance is  broken by the convection
term $ a \cdot{d \over dx}$ and
different exponential growth rates are found at $x \to + \infty$ 
and $x \to - \infty$.
On qualitative grounds, the asymptotic behavior is 
$\psi(\pm L) \approx exp[(\pm \cdot a - <z_{r}(E)>) \cdot
L]\,\,(L>>1)\,\,\,$ and delocalization occurs  when $<z_{r}(E)>-a=0$.

If one gauges away the IMF from the Schrodinger equation,
 Eq. \ref{A1} is restored, but the 
 boundary conditions are changed:  e.g., in the case of a double-sided 
problem, from 
$\psi(-L)= \alpha_{-},\, \psi(L)=\alpha_{+}$, 
one  has $\psi(-L)= exp(a \cdot L) \cdot \alpha_{-},\,
\psi(L)=exp(-a \cdot L) \cdot \alpha_{+}$.

We are not going to study the spectrum,  
but rather the vanishing of the ``effective'' Lyapunov exponent 
$<z_{r}(E)>-a$.
In order to do that it is sufficient to  study  the steady
state distribution  $P(z)=P(z_{r},z_{i})$ of Eq. \ref{A1} with $a=0$,
obtain $<z(E)>$ and from it the contour lines of $<z_{r}(E)>$.   
Notice that from the analysis a new object emerges, $<z_{i}(E)>$,
a sort of wavenumber of delocalized solutions.

In  the search for steady
state, we adopt 
an adiabatic approach.
This is possible with our model
potential, due to a property  of Eq. \ref{A1} 
with $V(x)=0$.
Let us  discuss this property first. 
 
Equation \ref{A1} has two time-independent
solutions (respectively stable and unstable critical point). 
If we denote with $\epsilon=\epsilon_{r} + i \cdot \epsilon_{i}$ one of the 
square roots of  $E$, the critical points are 
$z_{s}=\epsilon_{i}-i \cdot \epsilon_{r}$ and
$z_{u}=-\epsilon_{i}+i \cdot \epsilon_{r}$,( the suffixes $s,u$
meaning stable and unstable respectively, 
with $\epsilon_{i}>0$).
It is easily verified that the Equation can be explicitly solved,
for arbitrary initial conditions. After some algebra one obtains
the announced property: the solutions satisfy  an  
exact relation of the form:
\begin{eqnarray}
\label{A2}
{|z(x)|^2 \over |\epsilon|^2} &=& { {cosh(2B)-cos(2A)} \over
{cosh(2B)+cos(2A)}}\\
A &=& \epsilon_{r} \cdot (x-x_{0})+\gamma_{r}\nonumber\\
B &=& \epsilon_{i} \cdot (x-x_{0})+\gamma_{i}\nonumber
\end{eqnarray}
where $\gamma_{r}$ and $\gamma_{i}$ and $x_{0}$ are constants.

As a model potential, we
take
$V(x)$  two-valued, piecewise constant, $V(x)=(V_{0},\,V_{1})$. 
The
lengths of the $V_{l}$ intervals are two independent random variables 
with distributions $ Q_{l}(x)={1 \over c_{l}} \cdot exp(- c_{l} \cdot
x),\,\,\,(l_{l}={1 \over c_{l}}) $.

In order to write the analog of Equation \ref{A1}, 
we first introduce an
independent stochastic
variable $s(x)$, which assumes the values  $ 0,\,1$ with distributions 
$Q_{0}(x),\,\,Q_{1}(x)$.
 The probability
$P_{0}(x),\,P_{1}(x)$  of $s(x)$ satisfies
the rate equation:
\begin{eqnarray}
{dP_{0} \over dx} &=&c_{1} \cdot P_{1} -c_{0} \cdot P_{0}\\
{dP_{1} \over dx} &=&c_{0} \cdot P_{0} -c_{1} \cdot P_{1},\nonumber
\end{eqnarray}   
since $s(x)$ stays on the l-th channel an
average ``time''${1 \over c_{l}}$.

The equation we looked for is then:
\begin{equation}
\label{A3}
{dz \over dx} = -(z^2 + E) + V_{0} + s(x) \cdot (V_{1} - V_{0}),\\
\end{equation}
and, in components:
\begin{eqnarray}
\label{A4}
{dz_{r} \over dx} &=& -z_{r}^{2} + z_{i}^{2} - E_{r}+ 
V_{0} + s(x) \cdot (V_{1} - V_{0})\\
{dz_{i} \over dx} &=& -2 \cdot z_{r} \cdot z_{i} - E_{i}\nonumber
\end{eqnarray}
As is well known, from Eq.\ref{A1}, if $V(x)$ is gaussian,
one gets a Fokker-Plank equation for the probability.
Equation \ref{A3} is no longer of Langevin type, but one 
can show  that the probability 
distribution still satisfies a differential equation. 
The details of the derivation
can be found in Ref. \cite{LGP} and in the original paper \cite{BP}. 
In terms of the probability  $P(s,z;x)={P_{0}(z,x),P_{1}(z,x)} ;
\,\,(P_{0}(z;x)=P(s=0,z;x),\,\,P_{1}(z;x)=P(s=1,z;x)$ one has:
\begin{eqnarray}
\label{A5}
{dP_{0} \over dx} &=& (-{d \over dz_{r}} F_{r}^{(0)} 
-{d \over dz_{i}} F_{i}^{(0)}) P_{0} + c_{1} \cdot P_{1}-  c_{0} \cdot
P_{0}\nonumber\\
{dP_{1} \over dx} &=& (-{d \over dz_{r}} F_{r}^{(1)}  
-{d \over dz_{i}} F_{i}^{(1)}) P_{1} - c_{1} \cdot P_{1}+  c_{0} \cdot
P_{0}\nonumber\\ 
F_{r}^{(0)}&=& -z_{r}^{2} + z_{i}^{2} - E_{r}+ V_{0}\\
F_{r}^{(1)}&=& -z_{r}^{2} + z_{i}^{2} - E_{r}+ V_{1}\nonumber\\
F_{i}^{(0)}&=&F_{i}^{(1)}= -2 \cdot z_{r} \cdot z_{i} - E_{i}\nonumber
\end{eqnarray}
We  finally state our adiabatic
approximation. 
We assume that the
coefficients $c_{0}$ and $c_{1}$ are very small with respect to the
characteristic ``frequencies'' of the two deterministic evolutions 
$z^{(0)}(x)$ and $z^{(1)}(x)$,
respectively associated with $E^{(0)}=E_{r}-V_{0} +i \cdot E_{i}$ and
$E^{(1)}=E_{r}-V_{1} +i \cdot E_{i}$.
Such frequencies, 
as one  easily realizes, 
 are the square roots $\epsilon^{(0)},\,\epsilon^{(1)}$ 
of $E^{(0)}$ and  $E^{(1)}$.

\section{Low energy case}
Let us examine the low   $E_{r}$ case. In both deterministic
systems we assume $|\epsilon_{r}|<|\epsilon_{i}|$.
This regime is out 
of the physical region since it implies
  $E_{r}<V_{l},\,(l=0,1)$;
in spite of that, its analysis 
 will be found to be useful in the sequel.
 
Upon averaging 
Eq.  \ref{A2} over the  variable B, which is fast varying 
with respect to A when $|\epsilon_{r}|<|\epsilon_{i}|$ , the factor
$ { {cosh(2B)-cos(2A)} \over{cosh(2B)+cos(2A)}}$ reduces to unity, and
we obtain:
\begin{equation}
\label{B1}
|z^{(l)}|^{2} = |\epsilon^{(l)}|^{2},\,(l=0,1)\\
\end{equation}
Each channel has a (metastable) attractor of circular 
shape, and
the stochastic  motion reduces to hops between such  concentric
orbits. The single channel deterministic dynamics 
is illustrated in Figs.1 and 2, where we show the vector field and the 
probability distribution for Eq. \ref{A3} when $s(x)$ is kept
constant.

We now proceed to determine the steady state of Eq. \ref{A5}.
We first eliminate the 
$z_{i}$-dependence 
from  $F^{(0)}_{r}$ and $F^{(1)}_{r}$ by means of  Eq. \ref{B1}, 
then average
over $z_{i}$, and obtain:
\begin{eqnarray}
\label{B2}
 {d \over dz_{r}}(-2 \cdot z_{r}^{2} +2 \cdot
(\epsilon^{(0)}_{i})^{2}) P_{0} 
&=& c_{1} \cdot P_{1}-  c_{0} \cdot P_{0}\\
 {d \over dz_{r}}(-2 \cdot z_{r}^{2} +2 \cdot
(\epsilon^{(1)}_{i})^{2}) P_{1} 
&=&- c_{1} \cdot P_{1}+  c_{0} \cdot P_{0}.\nonumber
\end{eqnarray}
Here $P_{0}=P_{0}(z_{r})$ and $P_{1}=P_{1}(z_{r})$, are  
  averaged over 
$z_{i}$. The sum of the two equations gives:
\begin{equation}
\label{B3}
(2 \cdot z_{r}^{2} -2 \cdot
(\epsilon^{(0)}_{i})^{2}) P_{0}+(2 \cdot z_{r}^{2} -2 \cdot
(\epsilon^{(1)}_{i})^{2}) P_{1} = K,\\
\end{equation}
where $K$ is a constant.

It is readily verified that only with $K=0$  the positivity 
of the $P$'s is preserved.  
When $|z_{r}|$ is large enough the 
flow is in the negative $z_{r}$ direction on both channels;  
it turns positive
 for $|z_{r}|<|\epsilon_{i}^{(l)}|\,\,(l=0,1)$ 
respectively.
 
Let us 
assume, without loss of generality,
$0<\epsilon_{i}^{(1)}<\epsilon_{i}^{(0)}$.
A simple inspection of the hopping between channels 
shows that the particle gets trapped   
in the interval $\epsilon_{i}^{(1)}<z_{r}<\epsilon_{i}^{(0)}$;
the probability  is  zero 
outside.
Notice that upon including an indeterminacy in the radii of the
two circular attractors 
we would get a non zero probability flow.

Apart from the normalization factor $H$, the solutions have the form:
\begin{eqnarray}
\label{B4}
P_{0}(z_{r}) &=& H^{-1}{1 \over |z_{r}^{2} - (\epsilon_{i}^{(0)})^{2})|}
\cdot \psi(z_{r})\nonumber\\
P_{1}(z_{r}) &=& H^{-1}{1 \over |z_{r}^{2} - (\epsilon_{i}^{(1)})^{2})|}
\cdot \psi(z_{r})\nonumber\\
\psi(z_{r}) &=& \big[ |{{z_{r} - \epsilon^{(0)}_{i}} 
\over{z_{r} - \epsilon^{(0)}_{i}}}| \big]^{\mu_{0}} \cdot
\big[ |{{z_{r} - \epsilon^{(1)}_{i}} 
\over{z_{r} - \epsilon^{(1)}_{i}}}|\big]^{\mu_{1}}\\
\mu_{l} &=& {c_{l} \over {4 \cdot \epsilon^{(l)}_{i}}}\nonumber\
\end{eqnarray}
The function $P_{l}$ has an integrable power law singularity at 
the $l$-th channel's 
stable point, and a zero at the opposite channel's  stable point.
If we denote with $p_{0}$ and $p_{1}$ the integrals of 
 $P_{0}(z_{r})$ and $P_{1}(z_{r})$, they must satisfy the 
global equilibrium condition:
\begin{equation}
\label{B5}
c_{0} \cdot p_{0} =c_{1} \cdot p_{1}.
\end{equation}
We determined the solution
 in the region $c_{0}+c_{1}=1$ 
under this condition.  
The  power exponent of the  distribution 
is a function of 
the ratio ${ c_{0} \over c_{1}}$, i.e. it is shaped by the average 
``residence times'' over the two
channels.
The mean value $<z_{r}>$ is increasing with the  ratio
$\rho= {\epsilon_{i}^{(0)} \over \epsilon_{i}^{(1)}}$, as shown in Fig.5.
 Since larger ratios
imply larger values of $|V_{0}-V_{1}|$, this simply means    
that the Lyapunov exponent increases with disorder, as it should be.

Any boundary condition $z_{r}(L)= \overline{ z_{L}}$ 
with $\overline{ z_{L}}$ lying outside the interval
$\epsilon_{i}^{(1)}<z_{r}<\epsilon_{i}^{(0)}$
cannot be fulfilled:  the density of states must then be
zero in this regime, as anticipated.
  
\section{High energy case}
We analyze now the regime $E_{r}>>0$, which corresponds to the condition
$ |\epsilon_{i}| <
|\epsilon_{r}|$ for both channels. To simplify the notation, unless
when strictly necessary, in the next formulas we drop the channel index
$(l)$. 
  Contrary to the former case, 
in Eq. \ref{A2} the variable  A is now fast varying
with respect to B. 
The average over A gives:
\begin{equation}
{ |z|^{2} \over |\epsilon|^{2}} = -1 + { { |z|^{2} +
|\epsilon|^{2}} \over {|\epsilon_{r} \cdot z_{i} -
\epsilon_{i} \cdot z_{r}|}},
\end{equation}
from which:
\begin{equation}
\label{C1}
|\epsilon|^{2}= |\epsilon_{r} \cdot z_{i} -
\epsilon_{i} \cdot z_{r}|.\nonumber
\end{equation}
In a single channel, the metastable attractor is   a couple of
parallel lines: one through the stable, the other through the unstable
critical point. Furthermore, the lines are orthogonal to the segment 
connecting such points.
It appears that, in going from the low to the high energy limit, the 
attractor undergoes dilatation in the direction of such lines.
This effect can already be seen in Fig.2, referring to the 
low energy regime.

 An indeterminacy arises, about the line  
currently occupied by the ``particle'':  
\begin{equation}
\label{C11}
 z_{i}={\epsilon_{i} \over \epsilon_{r}} \cdot z_{r} 
\pm {|\epsilon|^{2} \over \epsilon_{r}}
\end{equation}
The single channel situation is  illustrated
in Figs.3 and 4, where we display the vector field and the probability 
distribution for Eq. \ref{A3} in the deterministic case.

The stochastic motion involves the union of two
couples of parallel lines, with distinct slopes.
We proceed with our strategy, based on averaging Eq. \ref{A5} over
$z_{i}$.
Upon writing
$F_{r}^{(l)}\,(l=0,\,1)$  
by means of Eq. \ref{C11}  
we have:
\begin{equation}
\label{C2}
F_{r}=-z_{r}^{2} + ({\epsilon_{i} \over \epsilon_{r}} \cdot z_{r})^{2}
\pm 2 \cdot \epsilon_{i} \cdot z_{r} + (\epsilon_{i})^{2},
\end{equation}
where the dependence on $E$ and $V_{l}$ is written in terms of 
 $\epsilon$ and
  the following approximation is made: 
\begin{equation}
z_{i}={\epsilon_{i} \over \epsilon_{r}} \cdot z_{r} \pm 
(\epsilon_{r}+{(\epsilon_{i})^{2} \over \epsilon_{r}}) \approx 
{\epsilon_{i} \over \epsilon_{r}} \cdot z_{r} \pm \epsilon_{r}.
\end{equation}
 The force $F^{(l)}(z_{r})\,\,(l=0,1)$ acquires
 a fluctuating term, originated in the indeterminacy of Eq. \ref{C11};   
 the term is of the order 
$  2 \cdot \epsilon_{i}^{(l)} \cdot z_{r}$.
 The dynamics associated with this force has the form:
\begin{equation}
{dz_{r} \over dx}=-z_{r}^{2} + 
({\epsilon_{i} \over \epsilon_{r}} \cdot z_{r})^{2} + (\epsilon_{i})^{2}+
2 \cdot \epsilon_{i} \cdot z_{r} \cdot \eta(x) ,
\end{equation}
where $\eta(x)$ is a white noise. 

In summary, the mentioned indeterminacy adds to the stochasticity
of the hopping between channels $l=0,1$; the process is then 
described by the Fokker-Plank equation:
\begin{eqnarray}
\label{C3}
{dP_{0} \over dx} &=& {d \over dz_{r}}(- A^{(0)} + 
{d \over dz_{r}} B^{(0)}) P_{0} + c_{1} \cdot P_{1}-  c_{0} \cdot
P_{0}\nonumber\\
{dP_{1} \over dx} &=& {d \over dz_{r}}(- A^{(1)} + 
{d \over dz_{r}} B^{(1)}) P_{1} - c_{1} \cdot P_{1}+  c_{0} \cdot
P_{0}\nonumber\\ 
A^{(l)}&=& - (\alpha_{l} \cdot z_{r})^{2} + (\epsilon^{(l)}_{i})^{2}\\
B^{(l)}&=& D \cdot (\epsilon^{(l)}_{i})^{2} \cdot [(z_{r})^{2} +
(\xi^{(l)})^{2}]\nonumber\\
\alpha_{l}^{2}&=& 1 -({\epsilon_{i} \over \epsilon_{r}})^{2}\nonumber
\end{eqnarray}
where $D$ is the white noise  coefficient and $\xi^{(l)}$ is 
a regularizing parameter (the criteria used in fixing this 
parameter are illustrated in the caption of Fig.6).
 
 We evaluated numerically, by a perturbative procedure,
the steady state solution of Eq. \ref{C3}.  
To simplify things,
we  assumed $c_{0}=c_{1}=c<1$. 
When the channels are uncoupled $(c=0)$, the problem can be explicitly
solved
along the lines of the hermitean case (see, e.g., Ref. \cite{BIH}).
The two equations reduce to the form: 
$\mathcal{M}^{(l)}P_{l}={d \over dz_{r}}J_{l}(z_{r})=0$, where, with 
obvious notation, $\mathcal{M}^{(l)}$ is the second order operator 
acting on $P_{l}$.

The solutions correspond to a constant probability current $J_{l}$ and
here look like:
\begin{eqnarray}
\label{C4}
P_{l}(z_{r})&=&{J_{l} \over B_{l}(z_{r})} \int_{- \infty}^{z_{r}} dz'
exp [U_{l}(z_{r})-U_{l}(z')]\\
U_{l}(z)&=&(1/D) \big[{ (1 + \alpha^{2}_{l}) \over \epsilon_{i}^{l}}
 \cdot artg({z \over \alpha_{l}})- { \alpha^{2}_{l} \over 
(\epsilon_{i}^{(l)})^{2}} \cdot z \big]\nonumber.
\end{eqnarray}

Upon substituting the series expansion
$P_{l} = \sum_{k} c^{k} \cdot P^{(k)}_{l},\,(l=0,1)$
in  Eq. \ref{C3}, 
one  obtains an inhomogeneous equation 
for $P^{(k+1)}_{l}$, where $P^{(k)}_{l}$ is the source term.
The operator $\mathcal{M}^{(l)}$ is easily inverted, with
two integrals over $z_{r}$.
The integrals are performed numerically. In order to check for convergence,
the area difference between successive approximants
$P^{n}_{l} = \sum^{n}_{k} c^{k} \cdot P^{(k)}_{l}$ 
is computed. In the cases exibited here, we put $c=0.1$,
and find good convergence for $n>20$.

From the  probability distribution $P_{l}(z)$ 
we  then get $<z_{r}>=<z_{r}>_{0}+<z_{r}>_{1}$, with obvious 
notation.

In Fig.6 we plot the result $<z_{r}>$ as a function of $E_{r}$.
The asymptotic behavior in the hermitean case
is known: $<z_{r}> \approx (E_{r})^{-2},(E_{i}=0)\,E_{r}>>1$\, \cite{LGP}.
Our estimate, with $E_{i}=10.$ and $E_{r}$ in the range: 
$50.<E_{r}<120.\,, $ is of a much slower
decay; 
further work
is needed to clarify this point.

When $E_{r}$ is fixed, $<z_{r}>$ increases with $|E_{i}|$, as 
shown in Fig.7; we get $<z_{r}> \approx |E_{i}|^{\beta}$, 
$\beta \approx 2$  for large enough $|E_{i}|$.

The value of the wavenumber $<z_{i}(E)>$, emerging in the
delocalized regime, can be roughly
estimated from $<z_{r}>$  and by choosing, in Eq. \ref{C11}, 
 the line through the stable point.

It is easily verified that $<z_{i}(E)>$ is an odd function of $E_{i}$
while $<z_{r}(E)>$ is even;  complex conjugate
eigenvalues carry opposite wavenumbers.
In the hermitean case the two eigenvalues  merge , and the wavenumber, 
being a byproduct of broken parity, is zero for standing waves.
\section{Intermediate energies}
So far we studied  two extreme cases, but intermediate energies in strong
disorder can as well be treated; let us consider the following regime: 
$V_{1} << E_{r} << V_{0}$.
 
 The  deterministic
 solutions
satisfy 
Eqs. \ref{B1} and \ref{C1} respectively, i.e. channel (0) is in 
a ``low energy'' regime, and channel (1) in a ``high energy'' one.
 The metastable attractor 
$\mathcal{L}$ is the union of
 a circle and of a couple of parallel lines. The system for
$P_{0}$ and $P_{1}$ is an hybrid between Eqs. \ref{B2} and \ref{C3}.

 The first order operator of Eq. \ref{B2} 
is coupled with the
second order one of Eq. \ref{C3}: this makes 
an iterative procedure hardly convergent. 
Based on results from direct numerical
integration of Eq. \ref{A3}, 
we assign 
an indeterminacy $\delta$ to the circular orbit:
\begin{equation}
{ |z|^{2} \over | \epsilon|^{2}} = 1 + \delta. \nonumber
\end{equation}
   
 An estimate for $\delta$ can 
be extracted \cite{BM} from the deterministic flow (see 
the thickness of  the ring  in Fig.2). 
The steady state equation then involves  two second order operators:
\begin{eqnarray}
\label{D1}
 {d \over dz_{r}}[ F^{(0)}_{r} - 
{d \over dz_{r}}\mathcal{D}^{0}  ] P_{0} 
&=& c_{1} \cdot P_{1}-  c_{0} \cdot P_{0}\nonumber\\
 {d \over dz_{r}}[ A^{(1)} - 
{d \over dz_{r}} B^{(1)}] P_{1}&=& - c_{1} \cdot P_{1}+  c_{0} \cdot
P_{0}\\
\mathcal{D}^{0}= (\delta \cdot | \epsilon^{0}|^{2})^{2}&&\nonumber
\end{eqnarray}
We proceed again by an iterative perturbation scheme.
At uncoupled channels  $(c=0)$ the ``low energy'' equation coincides 
now  with the well-known
 one (\cite{BIH}), holding in the hermitean case. 
Its solution carries a non zero probability current: in the present
context,
it comes from the spread of the circular attractor.
The ``high energy'' integral was already exhibited in the previous
section.

In Fig.9 we show the result, the surface $<z_{r}(E)>$. 
 The high energy regime surface 
is reported for comparison in Fig.8;
its contour lines in the $E$ plane 
are also shown in Fig.10. 
Qualitative agreement is found with previous numerical and analytical 
results. 
An estimate of $<z_{i}(E)>$ can be derived from the adiabatic
equations \ref{C11} and \ref{B1}.
  
\section{Conclusions}
We studied nonhermitean delocalization in the presence of disordered potentials
of Kronig-Penney type, where
the barriers have random length.
We translated  the hamiltonian eigenvalue problem into 
a stochastic equation for the logarithmic 
derivative of the wave function ($z={\psi' \over \psi}$).
From the steady state distribution of $z$ we determined the Lyapunov
exponent.

Our approach, valid 
under suitable 
adiabatic conditions, describes the steady state
in terms of the real part of $z$. 
The imaginary part of $<z>$, an effective wavenumber, 
can be recovered 
by means of the adiabatic formulae \ref{B1},\ref{C11}.

We analyzed two regimes: 
a)intermediate energies ($V_{1}<<E_{r}<<V_{0}$); 
b) high energies($V_{1}\,\,,V_{0}<<E_{r}$).
The behavior of the system in other
regimes
is currently under investigation \cite{BM}.

 In case (a), the particle hops in the $z$ plane between
a strip and a circle, the strip carrying a probability current.
The circle, which acts as a trap  for the particle, is absent in case (b), 
where the 'particle' hops between two strips, with different slopes.
Two currents coalesce in this case.

We obtained that $<z_{r}>$ is always decreasing with
$E_{r}$: at higher
energies the wave functions have larger localization length.

At fixed $<E_{r}>$,  $<z_{r}>$ increases  with $|E_{i}|$: this means
that dissipation (associated with $E_{i}$) enhances the localization.
The effective wavenumber $<z_{i}>$ increases with $E_{r}$, as 
one would expect with normal dispersion.

The curve $<z_{r}(E)>-a=0$,  $a$ being the convection
coefficient, determines the mobility edge in the complex 
energy plane.

We finally add some comments on the density of states (DOS).
Most analytical results on the DOS in the presence of an 
IMF, possibly with the single exception of the semiclassical
analysis by Silvestrov \cite{S}, are concerned with the discrete 
case. Disorder averaging of the resolvent operator can indeed be
performed in various discrete models, thus obtaining the DOS in 
explicit form \cite{BSB},\cite{BZ},\cite{FZ},\cite{GK}.
	
 The Langevin approach 
gives the DOS for the hermitean hamiltonian in the
continuum.  
As a preliminary step in that procedure, 
one writes the definition of the DOS
in terms of the $z$ variable: this simply amounts to 
requiring that $z(x,E)$ must fulfill the boundary condition: 
$z(L,E)=\overline{z_{L}}$.  
This condition is all is needed to determine the DOS, as long as
the mapping from $E$ to $z(L,E)$ is invertible, which is 
precisely the case in the hermitean problem.

Whether anything similar holds in the complex case is, as
far as we know, an open question.  

\newpage
\begin{figure}
\includegraphics[width=12 cm] {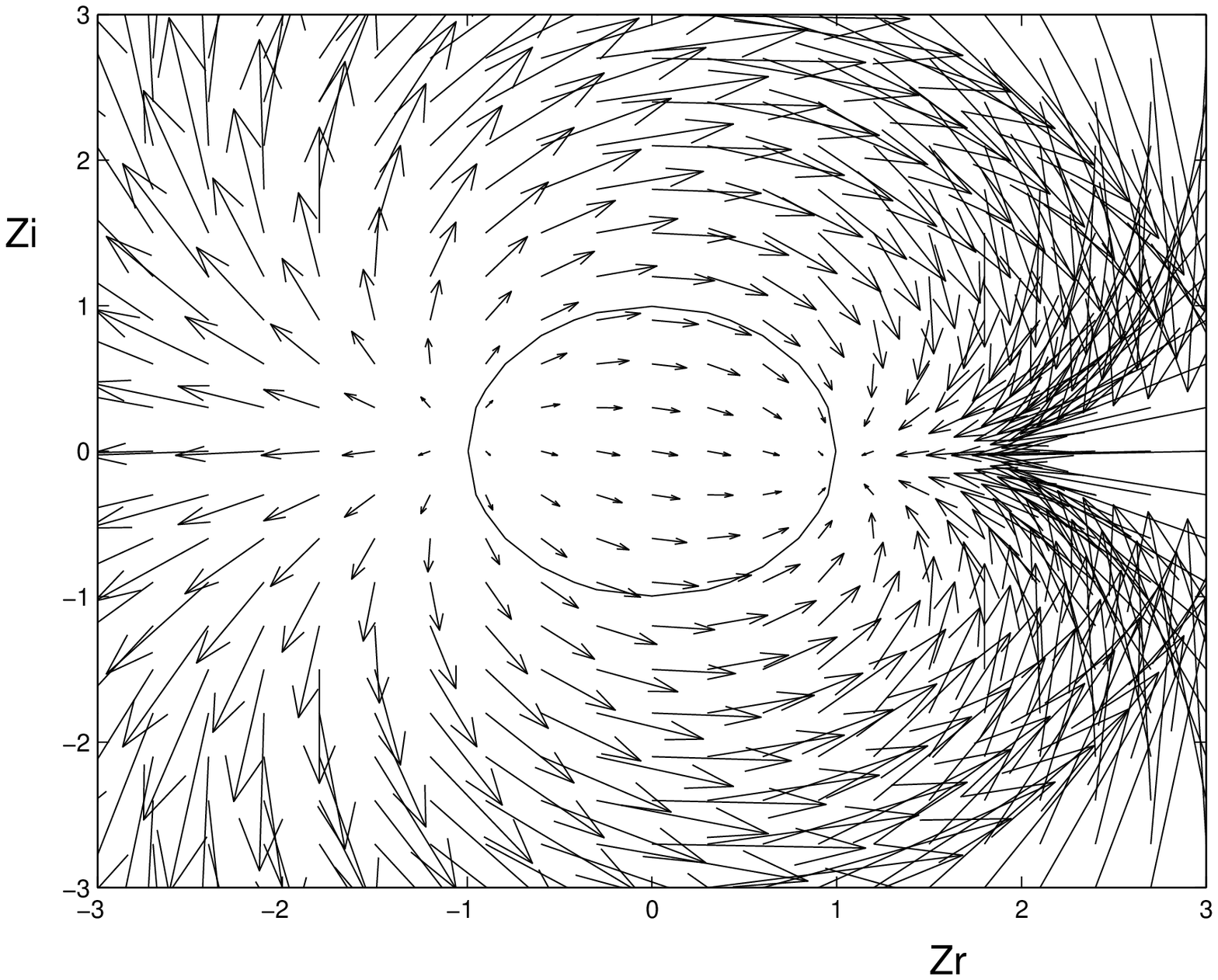}\\
Fig.1:Vector field $(F_{r},F_{i})$ in the $z$ plane for the
Equation \ref{A3} at fixed $s(x)$ (deterministic), 
in the low energy regime.
The variables $z_{r},\,z_{i}$ are in arbitrary units of $(length)^{-1}$.
The superimposed circle corresponds to the metastable attractor
discussed in the text (see Eq. \ref{B1}). The two critical
points characterized by the condition $F_{r}=0,F_{i}=0$ belong 
to the circle.\\
\end{figure}
\begin{figure} 
\includegraphics[width=12 cm] {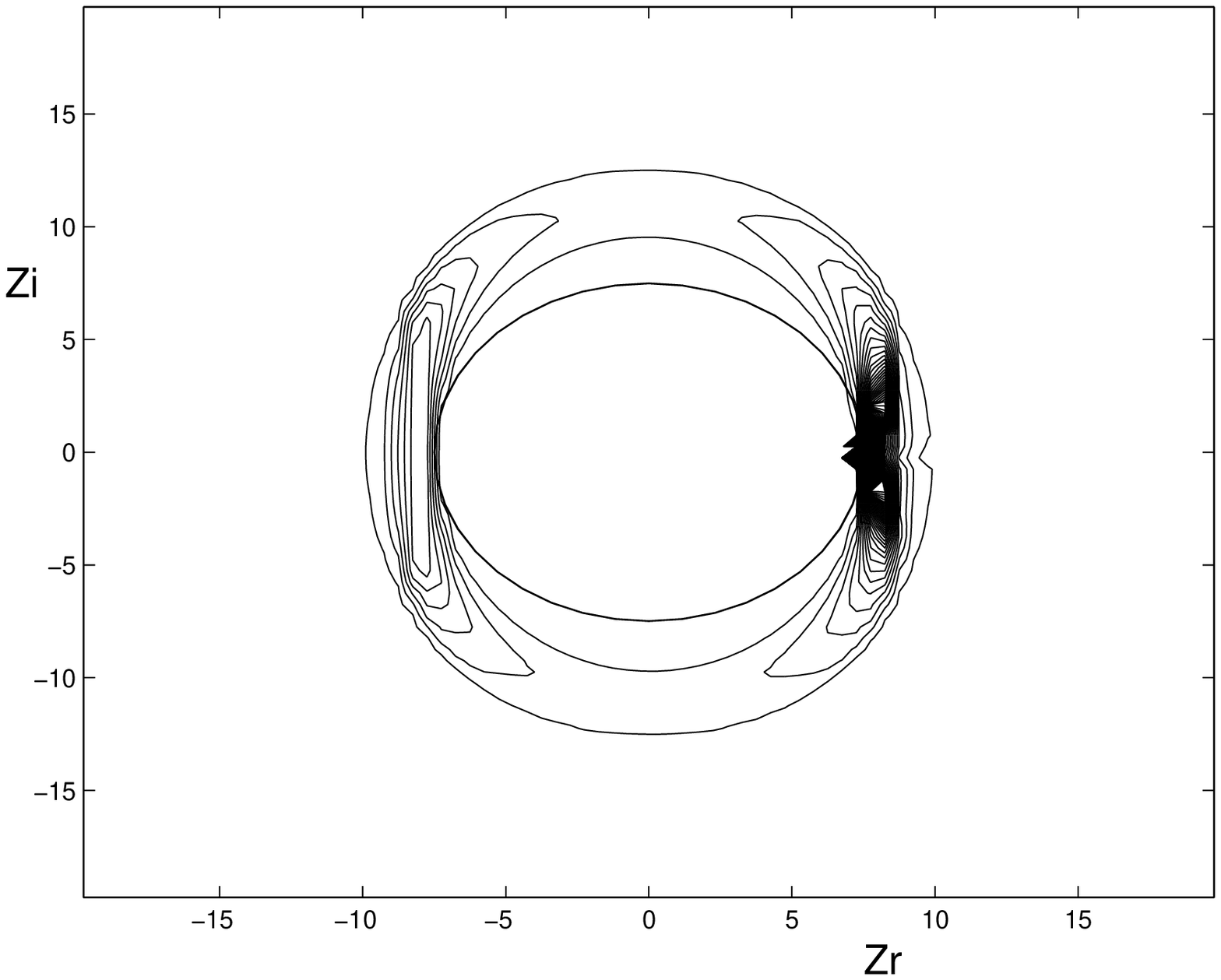}\\
Fig.2:Numerical integration of the Equation \ref{A3} with $s(x)=0$ in the 
low energy regime: contour lines
of the probability distribution at an intermediate time. 
The variables $z_{r},\,z_{i}$ are in arbitrary units of $(length)^{-1}$.
The distribution is obtained from 2000 random initial conditions.
The inner
circle is superimposed: it corresponds to the metastable
attractor, 
as calculated 
from Eq. \ref{B1}, with the values
$E_{r}-V_{0}=-56,E_{i}=-1.35$ used in the numerical integration.
The variables $E_{r},\,E_{i}$ are in arbitrary units of $(length)^{-2}$.
The unstable critical point is on the left of the ring; 
the peak at the stable point can be seen on the right.
The distribution  deviates from 
a purely circular behavior; it is  elongated in the aequator
direction,
if one takes the critical points as  the poles.
In fact going from the low to
 the high energy regime (see Eq.\ref{C11}) the  attractor  is
converted 
into a couple of lines. These lines are 
through the poles, and parallel 
to the aequator.\\
\end{figure}
\begin{figure}
\includegraphics[width=12 cm] {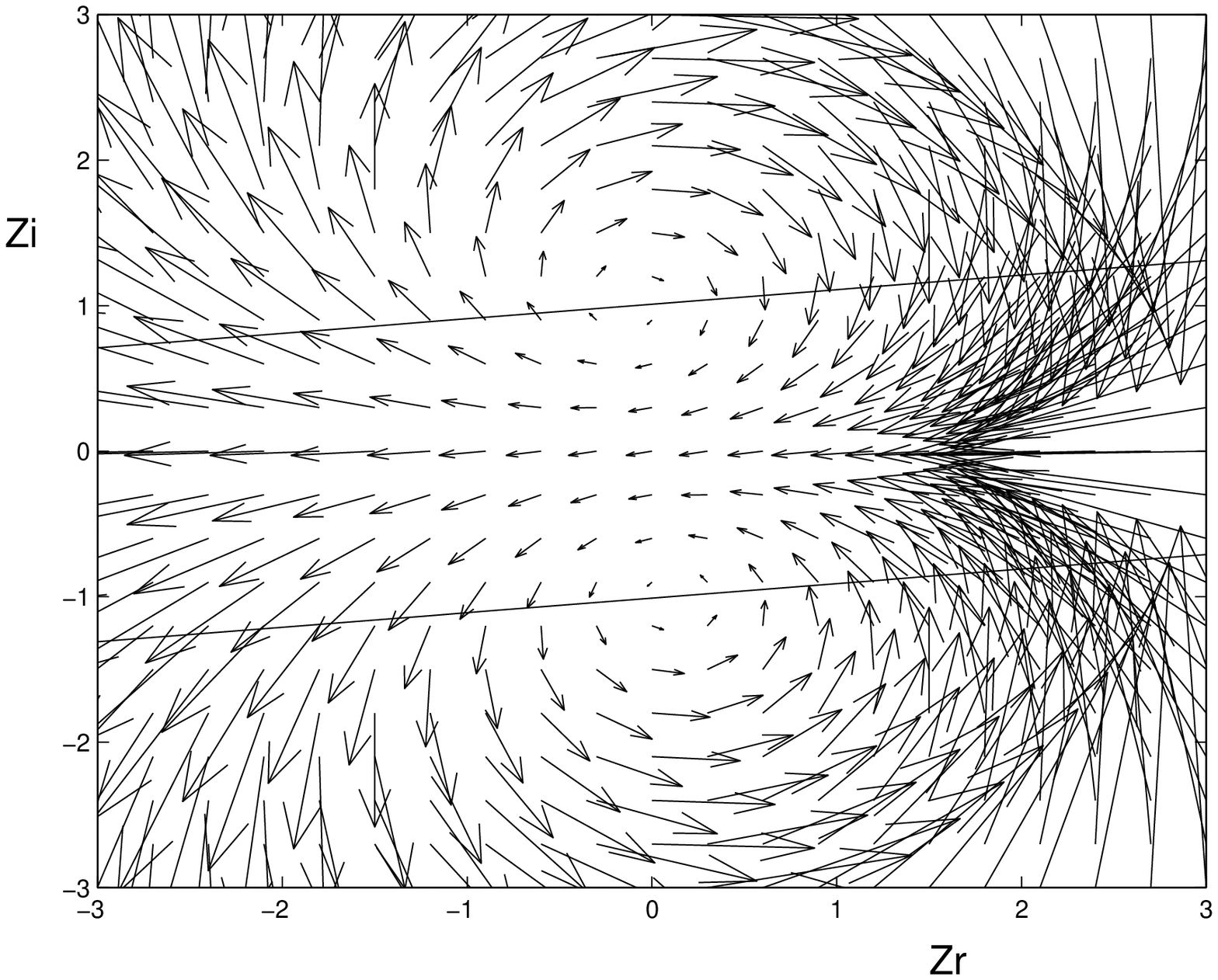}\\
Fig.3:Vector field $(F_{r},\,F_{i})$ of  Equation \ref{A3} 
with fixed $s(x)$ in the high energy
regime. 
The variables $z_{r},\,z_{i}$ are in arbitrary units of $(length)^{-1}$.
The two superimposed parallel lines through the critical points
$(F_{r}=0,\,F_{i}=0)$  correspond to
the metastable attractor (see Eq. \ref{C11}).
The lines are orthogonal to the segment connecting the critical points.\\
\end{figure}
\begin{figure} 
\includegraphics[width=12 cm] {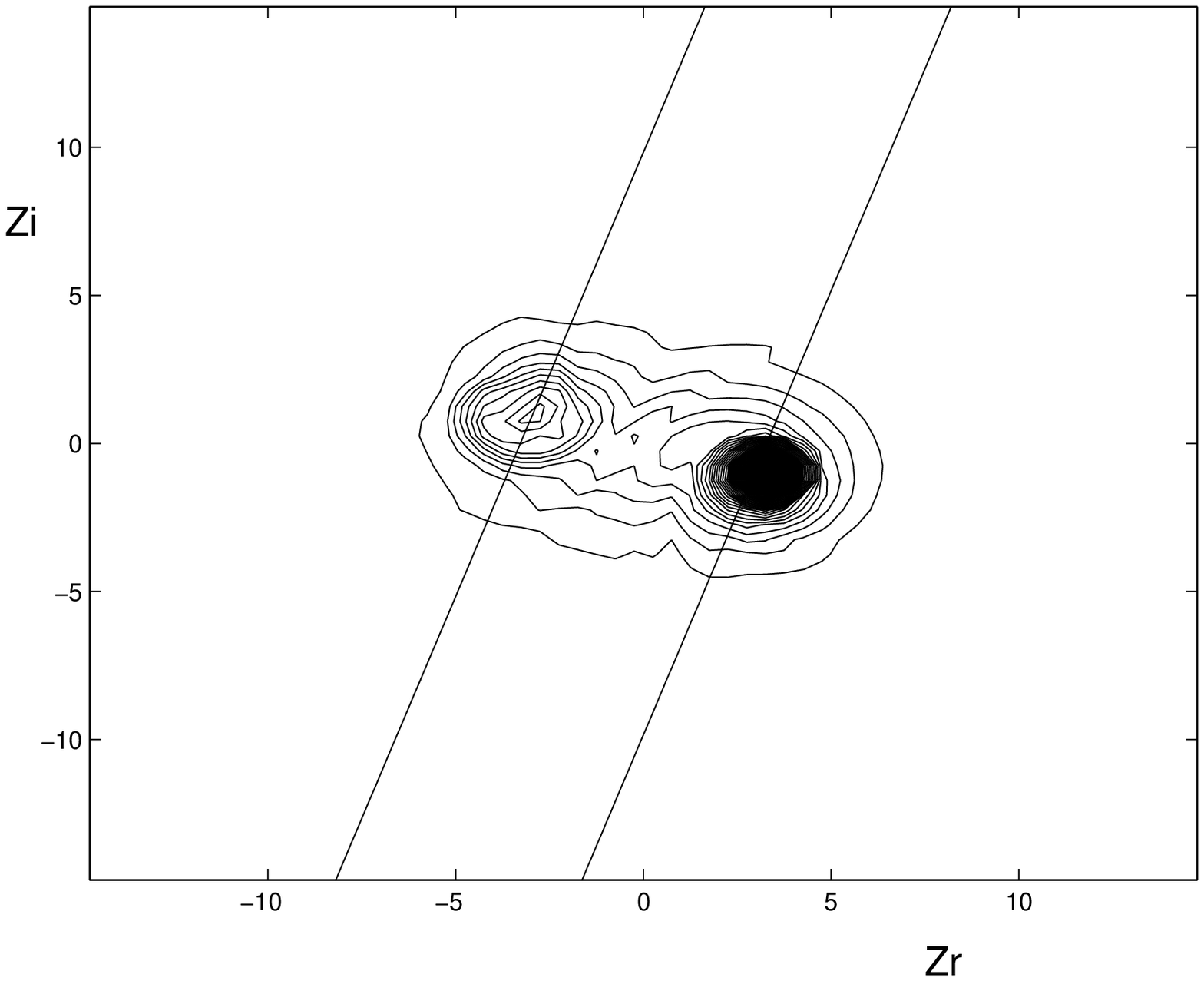}\\
Fig.4:Numerical integration of Eq. \ref{A3} with  
$s(x)=1$ 
in the 
high energy regime (random initial conditions): contour lines 
of the probability distribution 
at an intermediate time.
The variables $z_{r},\,z_{i}$ are in arbitrary units of $(length)^{-1}$.
The energy values are $E_{r}-V_{1}=10.35,\,E_{i}=7.5$,
in arbitrary units of $(length)^{-2}$.
The unstable critical point is in the upper left.
The metastable attractor (two parallel lines) has been superimposed;
with respect to Fig.3,  we are here at lower energies (the critical 
points are closer to the axis $z_{r}$).
One can notice, in particular around the unstable point, that the 
distribution is slightly elongated in the direction of the attractor.\\
\end{figure}
\begin{figure}  
\includegraphics[width=12 cm] {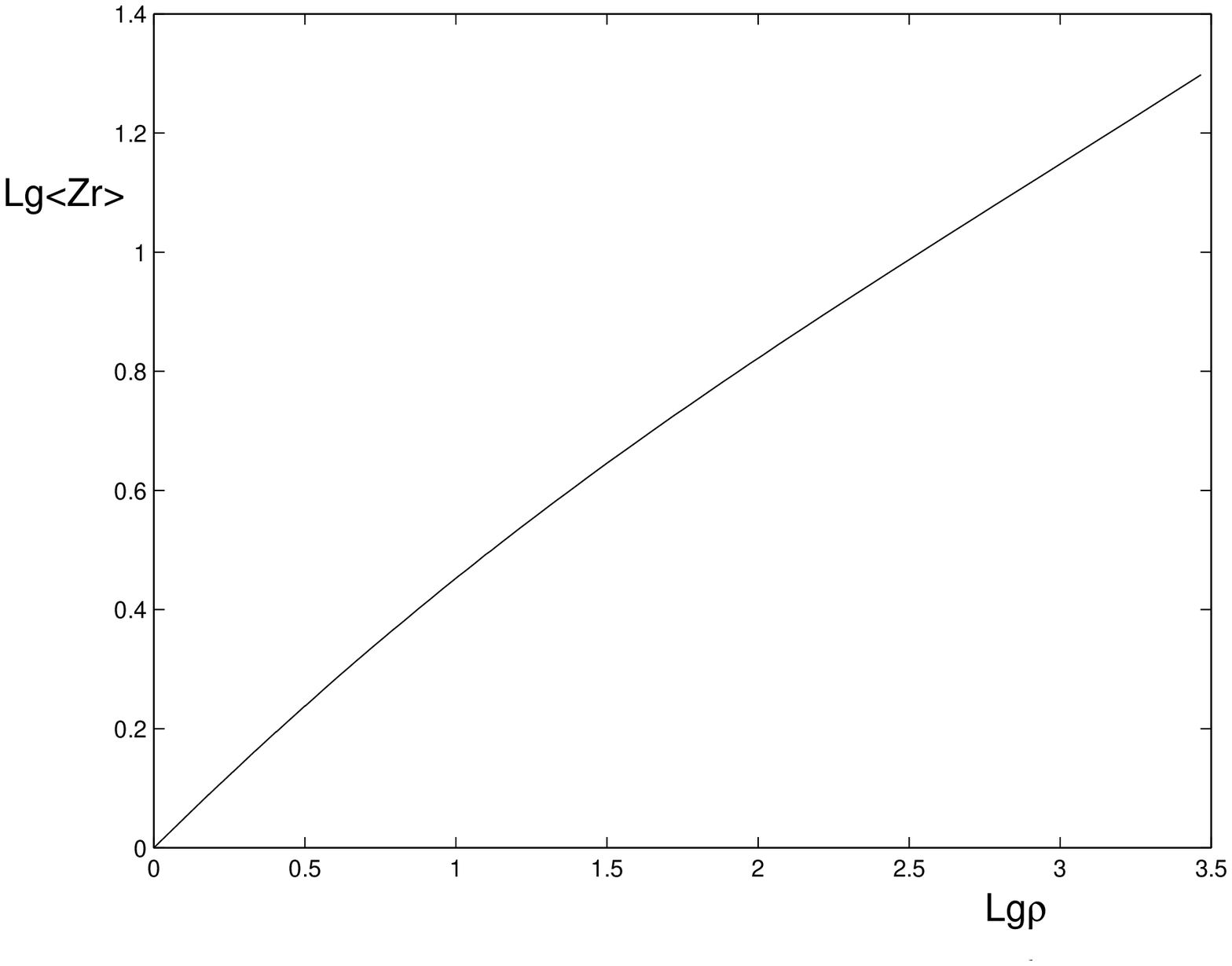}\\
Fig.5: Low energy regime; from the solution (Eq. \ref{B4}) of the 
steady state Equation \ref{B2} for the probability: 
log-log plot of $<z_{r}>$ versus 
$\rho={\epsilon_{i}^{(0)}\over \epsilon_{i}^{(1)}}$.
Higher $\rho$ means higher disorder (see text):
 the localization length decreases with $\rho$.\\
\end{figure}
\begin{figure}
\includegraphics[width=12 cm] {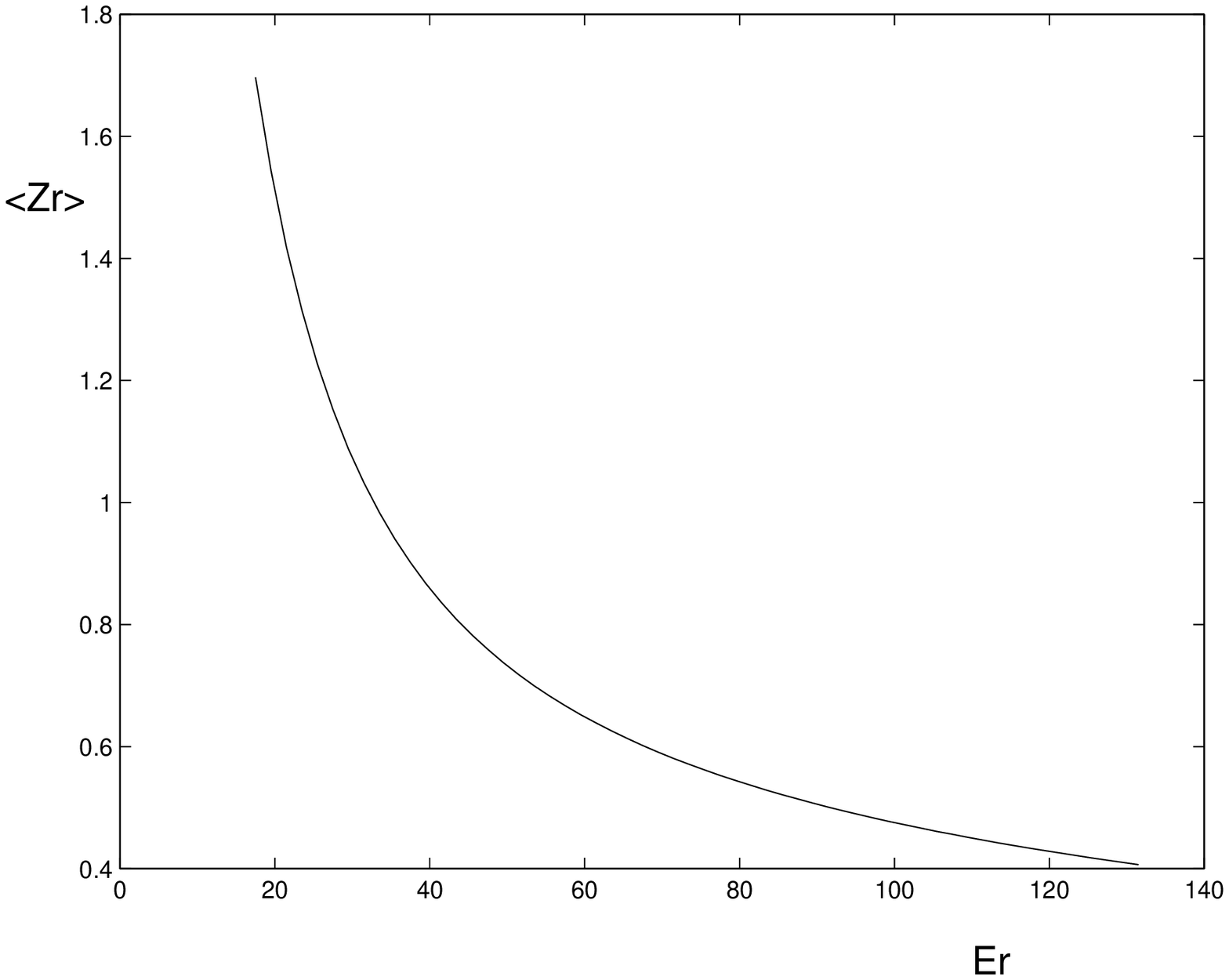}\\
Fig.6: High energy regime: $<z_{r}>$ versus 
$E_{r}$ at fixed $E_{i}$,\,$(E_{i}=10)$. 
The variable $z_{r}$ is in arbitrary units of $(length)^{-1}$,
and $E$ in arbitrary units of $(length)^{-2}$.
The origin of the energy 
axis corresponds to 
$E_{r}={1 \over 2} \cdot (V_{(0)}+ V_{(1)}) $;
\,$V_{(0)}- V_{(1)}=5.$
From numerical integration of the probability Equation\ref{C3} at steady
state, 
with\,$D=0.1$.
The regularization parameter is: $\xi^{(l)} =
\epsilon^{l}_{i}$. With this choice  $\xi^{(l)}$ has the same order of
magnitude
of $z_{r}$ at the stationary points:
  $z_{r} = \pm \epsilon^{(l)}_{i}$.\\
\end{figure}
\begin{figure} 
\includegraphics[width=12 cm] {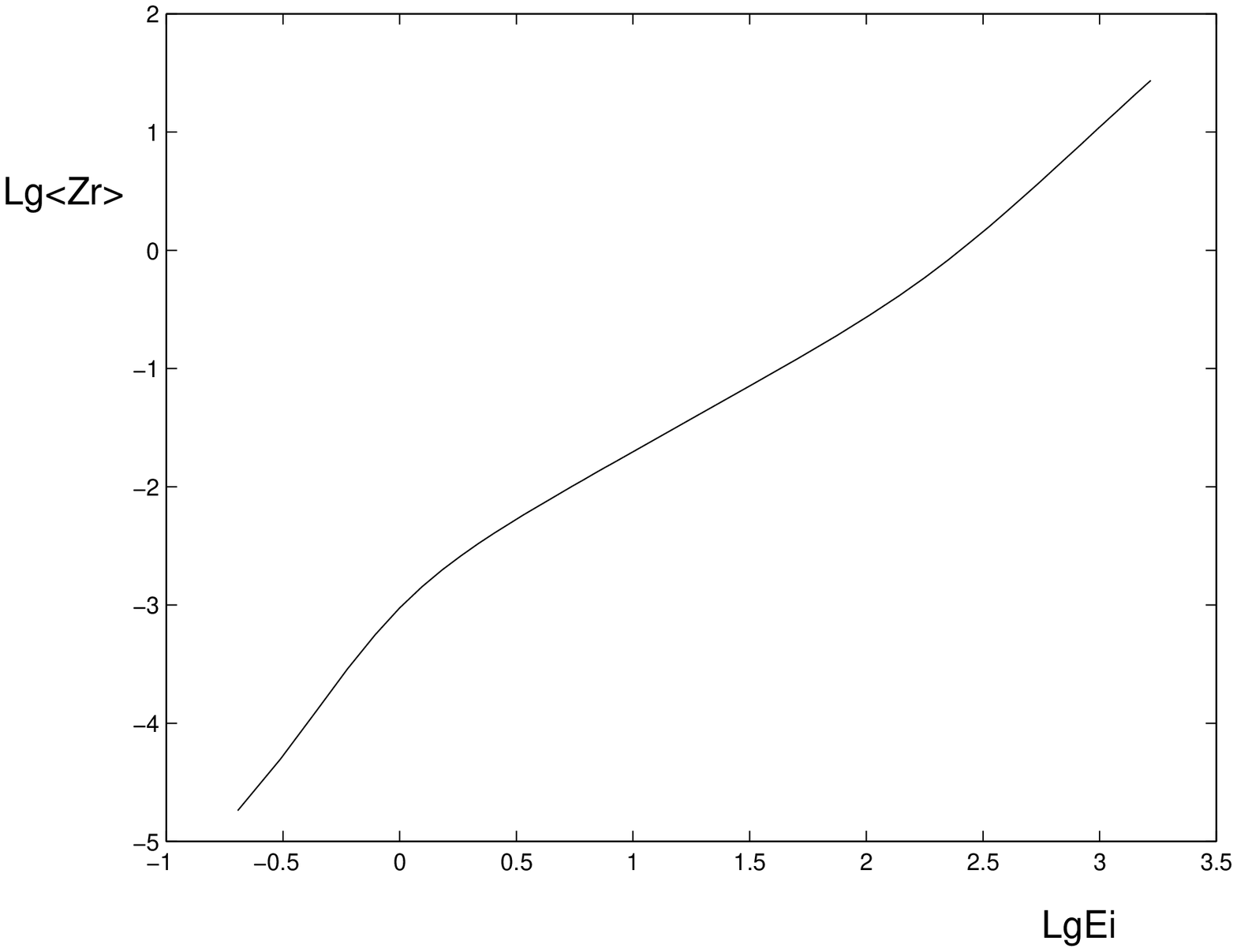}\\
Fig.7: High energy regime: log-log plot of $<z_{r}>$ versus $E_{i}$ at fixed 
$E_{r}$:$\,\, E_{r}-V_{(1)}=60.,\,
E_{r}-V_{(0)}=30.$
Same Equation and parameters as in Fig.6.\\
\end{figure}  
\begin{figure} 
\includegraphics[width=15  cm]{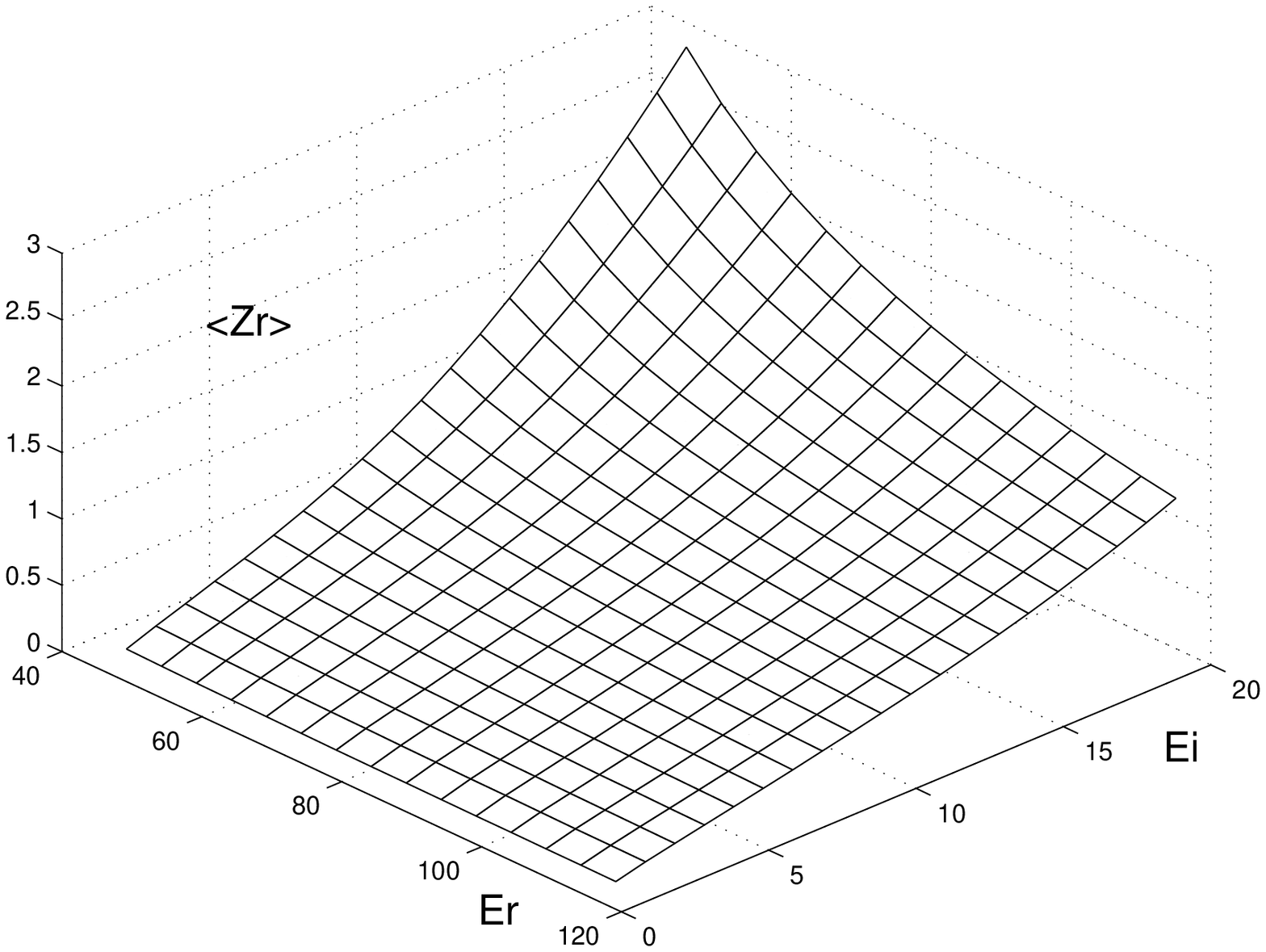}\\
Fig.8: High energy regime:global behavior of $<z_{r}>$ versus 
$E$;\,
$V_{(0)}-V_{(1)}=30.$
The variable $z_{r}$ is in arbitrary units of $(length)^{-1}$,
and $E$ in arbitrary units of $(length)^{-2}$.
From numerical integration of  Equation \ref{C3} for the steady 
state probability. 
Same parameters 
as in Figs. 6,7.\\
\end{figure}
\begin{figure}  
\includegraphics[width=15 cm]{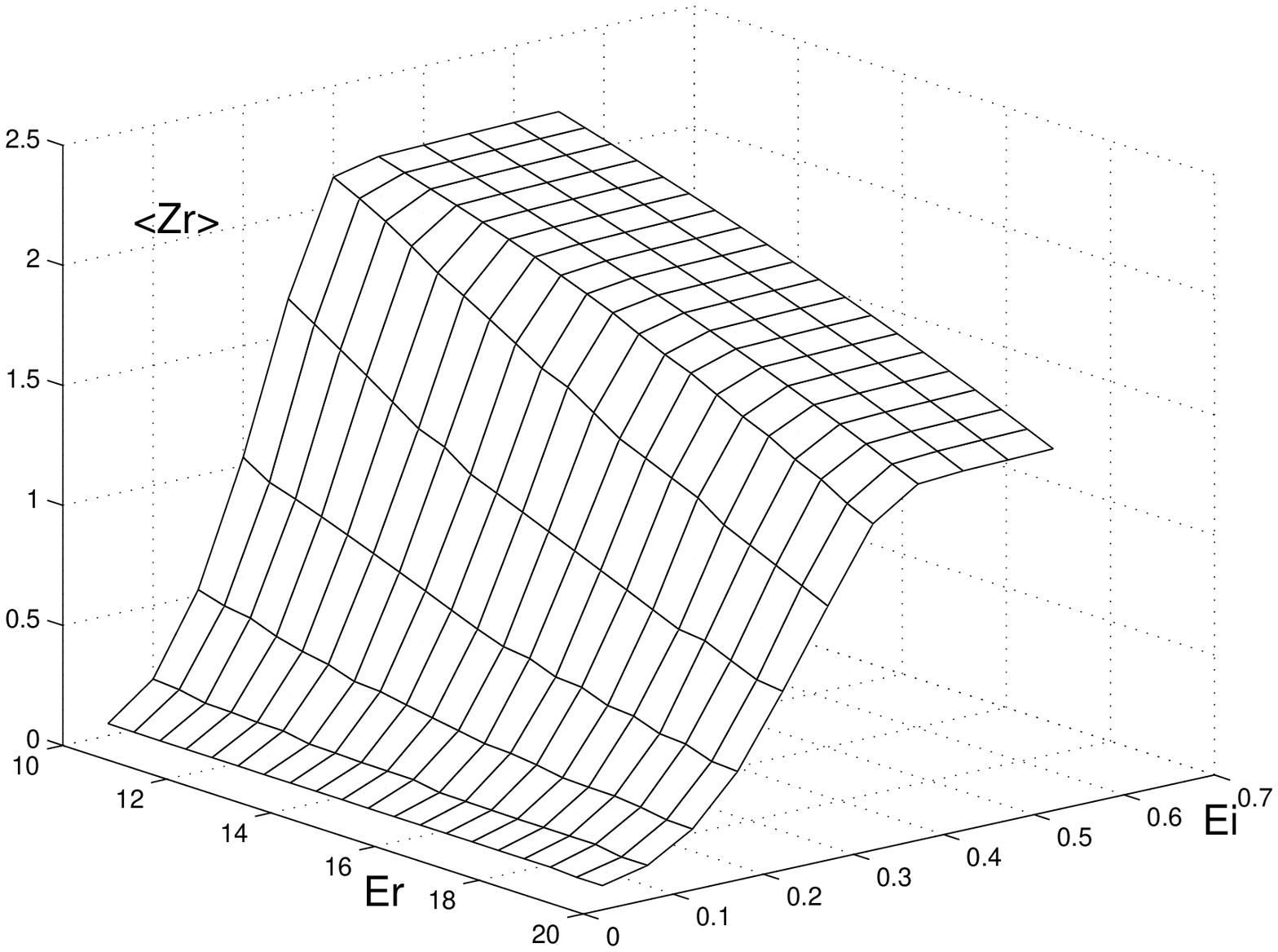}\\
Fig.9: Intermediate energy regime, from numerical integration of 
the  Equation \ref{D1} for the steady state probability: 
$<z_{r}>$ versus $E$,\,$D=0.1$,
$\delta=25.$,\,$V_{(0)}-V_{(1)}=30.$
The variable $z_{r}$ is in arbitrary units of $(length)^{-1}$,
and $E$ in arbitrary units of $(length)^{-2}$.   \\
\end{figure}
\begin{figure}  
\includegraphics[width=12 cm]{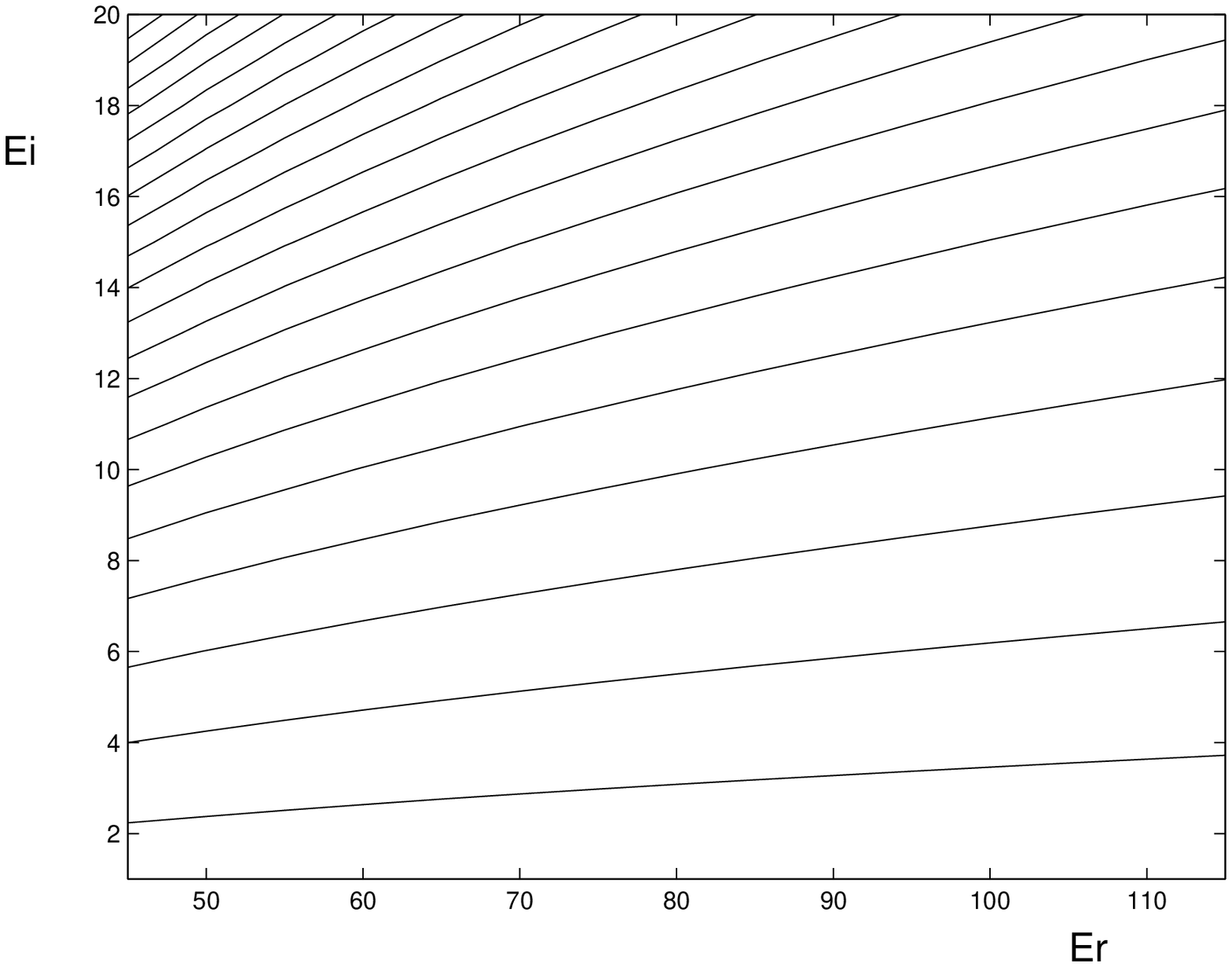}\\
Fig.10:High energy regime.Contour lines of the surface $<z_{r}(E)>$ 
plotted in Fig.8. 
 $E_{r},\,E_{i}$ are in arbitrary units of $(length)^{-2}$.
Each line
identifies a mobility edge, for a given convection 
parameter $a$;
 $a$ increases from bottom right to top left, in a range of values 
from 0.1 to 2.95, with step 0.15. \\          
 \end{figure}
\end{document}